\documentclass[conference]{IEEEtran}
\IEEEoverridecommandlockouts

\usepackage{cite}
\usepackage{amsmath,amssymb,amsfonts}
\usepackage{algorithmic}
\usepackage{graphicx}
\usepackage{textcomp}
\usepackage{microtype}
\usepackage{amsmath}
\usepackage{enumitem}

\usepackage{mathtools, nccmath}
\usepackage{lipsum}
\usepackage{graphicx}
\usepackage[caption=false]{subfig}
\usepackage{ifthen}
\usepackage{multirow}
\usepackage{comment}
\usepackage{eurosym,booktabs}
\usepackage{multicol, blindtext}
\usepackage{multicol, textcomp}
\usepackage{enumitem}
\usepackage{array}
\usepackage[utf8]{inputenc}
\usepackage[english]{babel}
\usepackage{soul}
\usepackage{url}

\usepackage{makecell}
\usepackage{wasysym}
\usepackage{xcolor}
\robustify\,
\usepackage{siunitx}
\def\BibTeX{{\rm B\kern-.05em{\sc i\kern-.025em b}\kern-.08em
    T\kern-.1667em\lower.7ex\hbox{E}\kern-.125emX}}

\newcommand{\NAME}{\mbox{\textsc{UMBRELLA}}\xspace}

\iftrue
    \newcommand{\todo}[1]{\color{red}TODO: #1\color{black}\xspace}

\else
    \newcommand{\todo}[1]{}

\fi

\newcommand{\zerodisplayskips}{%
  \setlength{\abovedisplayskip}{3pt}%
  \setlength{\belowdisplayskip}{3pt}%
  \setlength{\abovedisplayshortskip}{3pt}%
  \setlength{\belowdisplayshortskip}{3pt}}
\appto{\normalsize}{\zerodisplayskips}
\appto{\small}{\zerodisplayskips}
\appto{\footnotesize}{\zerodisplayskips}

\begin{document}
\bstctlcite{IEEEexample:BSTcontrol}

\title{Past, Present, Future: A Comprehensive Exploration of AI Use Cases in the \NAME IoT Testbed}

\author{\IEEEauthorblockN{
Peizheng Li\IEEEauthorrefmark{1},
Ioannis Mavromatis\IEEEauthorrefmark{2},
Aftab Khan\IEEEauthorrefmark{1}
}\\ 
\vspace{-3.00mm}
\IEEEauthorblockA{
\IEEEauthorrefmark{1} Bristol Research and Innovation Laboratory, Toshiba Europe Ltd., U.K.\\
\IEEEauthorrefmark{2}Digital Catapult, London, U.K.\\
Email: {\{Peizheng.Li, Aftab.Khan\}@toshiba-bril.com}, Ioannis.Mavromatis@digicatapult.org.uk
}}

\maketitle

\begin{abstract}
\NAME is a large-scale, open-access Internet of Things (IoT) ecosystem incorporating over 200 multi-sensor multi-wireless nodes, 20 collaborative robots, and edge-intelligence-enabled devices. This paper provides a guide to the implemented and prospective artificial intelligence (AI) capabilities of \NAME in real-world IoT systems. Four existing UMBRELLA applications are presented in detail: 1) An automated streetlight monitoring for detecting issues and triggering maintenance alerts; 2) A Digital twin of building environments providing enhanced air quality sensing with reduced cost; 3) A large-scale Federated Learning framework for reducing communication overhead; and 4) An intrusion detection for containerised applications identifying malicious activities. Additionally, the potential of \NAME is outlined for future smart city and multi-robot crowdsensing applications enhanced by semantic communications and multi-agent planning. Finally, to realise the above use-cases we discuss the need for a tailored MLOps platform to automate \NAME's model pipelines and establish trust.
\end{abstract}


\section{Introduction}
The Internet of Things (IoT) stands as a crucial research field, revolutionising industries, addressing societal challenges, and fostering innovation in connectivity, security, energy efficiency, and societal impact. Nowadays, the advancements in radio technologies, network protocols, and artificial intelligence (AI)/Machine learning (ML) bring more choices for designing an efficient IoT system with appropriate device selection, network architecture, and data processing capability.
Nevertheless, the flip side of this coin introduces more critical challenges in building an IoT system as people expect this IoT network to have enough flexibility for pre-simulation, reconfiguration, scalability control, and, most importantly, intelligence and automation capability~\cite{shafique2020internet}. 

For the existing IoT testbeds, considering their distinctive features, supported use cases, and major focuses, they can be classified into three main categories: wireless experimentation, robotics research, and AI-related activities\footnote{It should be noted this classification has no clear boundaries, as there is some overlap among the functionality of testbeds.}. Here, we list some of the operational and accessible platforms of each category and briefly introduce their objectives.
\begin{itemize}[leftmargin=*]
    \item \textbf{Wireless Experimentation}: In this class, examples like FIT IoT-Lab~\cite{adjih2015fit}, FlockLab 2 Testbed~\cite{trub2020flocklab}, and w-ILab~\cite{bouckaert2011w} provide public access, enabling Low-power WAN (LPWAN) experimentation. They are used primarily to test wireless networking protocols and algorithms.
    \item \textbf{Robotics Research}: Robotarium~\cite{pickem2017robotarium}, IRIS~\cite{tran2018intelligent}, and LivingLab~\cite{FraunhoferIML} are notable for robotics and swarm activities. Robotarium features miniature custom-designed robots, enhancing battery autonomy and reducing the testbed's footprint but limiting hardware capabilities. IRIS robots have fully programmable wireless IEEE 802.15.4~\cite{9144691} radio interfaces, yet position accuracy degrades in long experiments due to accumulated drift. FIT IoT-Lab and w-ILab are still under development, resulting in limited functionality.
    \item \textbf{AI/ML related}: For AI/ML, platforms like Gradient~\cite{Gradient} and OpenAI~\cite{openai} offer subscription-based access for evaluating AI/ML algorithms. Google Colab provides similar functionality for free. Despite powerful hardware, these platforms lack access to real-time generated data.
\end{itemize}

In general, the existing testbeds exhibit limited designed capabilities for handling only specific tasks, with short-supported use cases and a notable absence of access to real-time, real-world data, especially in AI/ML-enabled platforms. In this context, this paper introduces the unique work of \NAME\footnote{UMBRELLA - A living lab: \url{https://www.umbrellaiot.com/}} - Urban Multi Wireless Broadband and IoT Testing for Local Authority and Industrial Applications~\cite{mavromatis2024umbrella}. Functioning as a dynamic testing environment, \NAME allows users to emulate realistic scenarios, prototype solutions, and assess their IoT innovations. It provides foundational capabilities, tools, and infrastructure for testing physical and digital solutions across various use cases, including Smart City initiatives, IoT/Industrial IoT (IIoT), and AI/ML applications. This paper delves explicitly into the AI/ML use cases developed within the \NAME ecosystem, emphasising their validation and deployment in real-world scenarios. All use cases discussed in this paper are supported by a unified cloud-native platform, to enhance model reliability, thereby enabling their scalability, extensibility and robustness. Simultaneously, the paper envisions the potential AI/ML use cases for \NAME in complex real-world scenarios for construing trustable crowdsensing systems, and the importance of embracing  ML Operations (MLOps), outlining them as future developments.
\begin{figure*}
\centering
\includegraphics[width=0.85\linewidth]{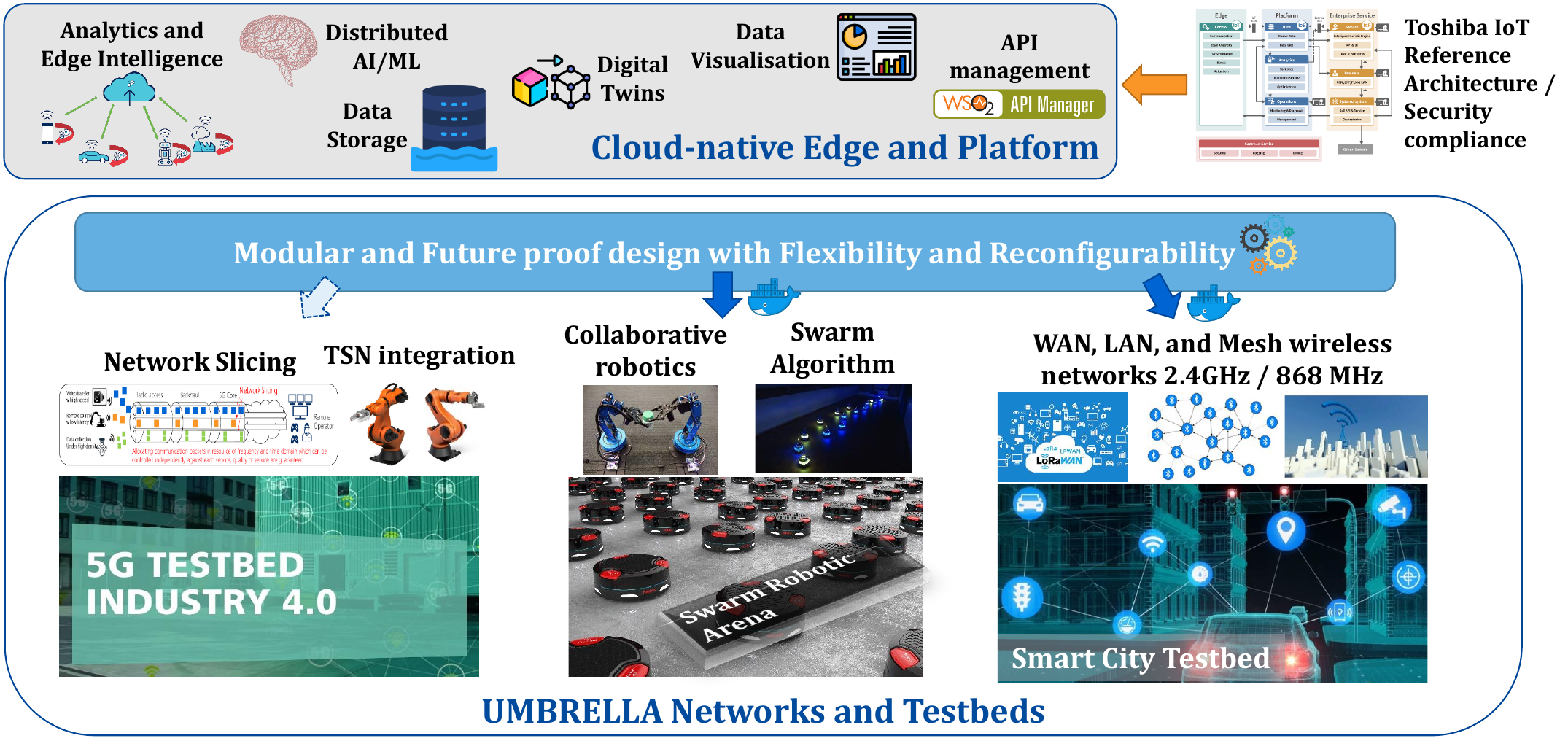}
\caption{\NAME SoS architecture overview with support of multiple sub-system testbeds.}
\label{fig:umbrella_pltform}
\vspace{-1.45mm}
\end{figure*}
The rest of the paper is organized as follows: Sec.~\ref{sec:testbed} introduces the \NAME ecosystem and provides details about its features, such as system components, hardware design, and the implementation method for in-\NAME applications. In Sec.~\ref{sec:existing AI usecase}, various AI/ML use cases demonstrated within \NAME are presented. This section includes both technical details and the results obtained. Sec.~\ref{sec:future AI} explores potential use cases scheduled to be implemented on the \NAME platform. Sec.~\ref{sec:MLOps} emphasises the importance of adopting full-scale MLOps for AI/ML use cases in \NAME. Finally, Sec.~\ref{sec:discussion and conclusion} concludes this paper.

\section{\NAME Testbed and Applications}
\label{sec:testbed}

\subsection{What is \NAME?}
\NAME~\cite{mavromatis2024umbrella} is a groundbreaking open-access IoT ecosystem deployed in the South Gloucestershire, U.K. region. Its primary goal is to stimulate innovation across diverse technological domains. Providing access to multiple testbed facilities, data, and computing resources, \NAME links specialised test environments and offers a System-of-Systems (SoS) approach to address real-world technological challenges. 
Comprising 200+ multi-sensor, multi-wireless interface nodes installed on public streetlights, a robotics arena with 20 mobile robots, and a 5G network-in-a-box solution, all controlled via a unified backend platform, \NAME ensures efficient management, control, and secure user access. 
\NAME's multi-domain architecture positions it as an ideal playground for IoT and IIoT innovation, embodying an open, sustainable testbed that actively guides future initiatives. With its unique blend of openness, heterogeneity, realism, and advanced tools, \NAME is dedicated to advancing technology research, development, and translation into tangible real-world progress. The \NAME SoS architecture overview supporting multiple sub-systems is illustrated in Fig.~\ref{fig:umbrella_pltform}. 

The guiding principle behind \NAME is that each existing system can operate independently, but the interoperation of various Constituent Systems (CSs) enables more complex scenario demonstrations. \NAME has contributed to multiple use cases since its deployment in Sep. 2021. It also facilitates multiple systems developed by Small and Medium-sized Enterprises (SMEs) and has been fully functional since then, collecting over six million sensing samples daily. Essentially, \NAME aims to foster a collaborative innovation ecosystem between users, industry, government, and academia to create economic and societal values.

\subsection{\NAME Hardware Design and the AI Capability}
The hardware design of \NAME places scalability and realism at the central stage; that is, the hardware should operate in realistic conditions and constraints, matching the requirements for the envisioned use cases. The hardware composition of \NAME sensor node and collaborative Robotics testbed is illustrated as follows.

\subsubsection{\NAME Sensors Node}
\begin{figure}[t]   
    \subfloat[\label{fig:node inner}]{
      \begin{minipage}[t]{0.45\linewidth}
        \centering 
        \includegraphics[width=1.7in]{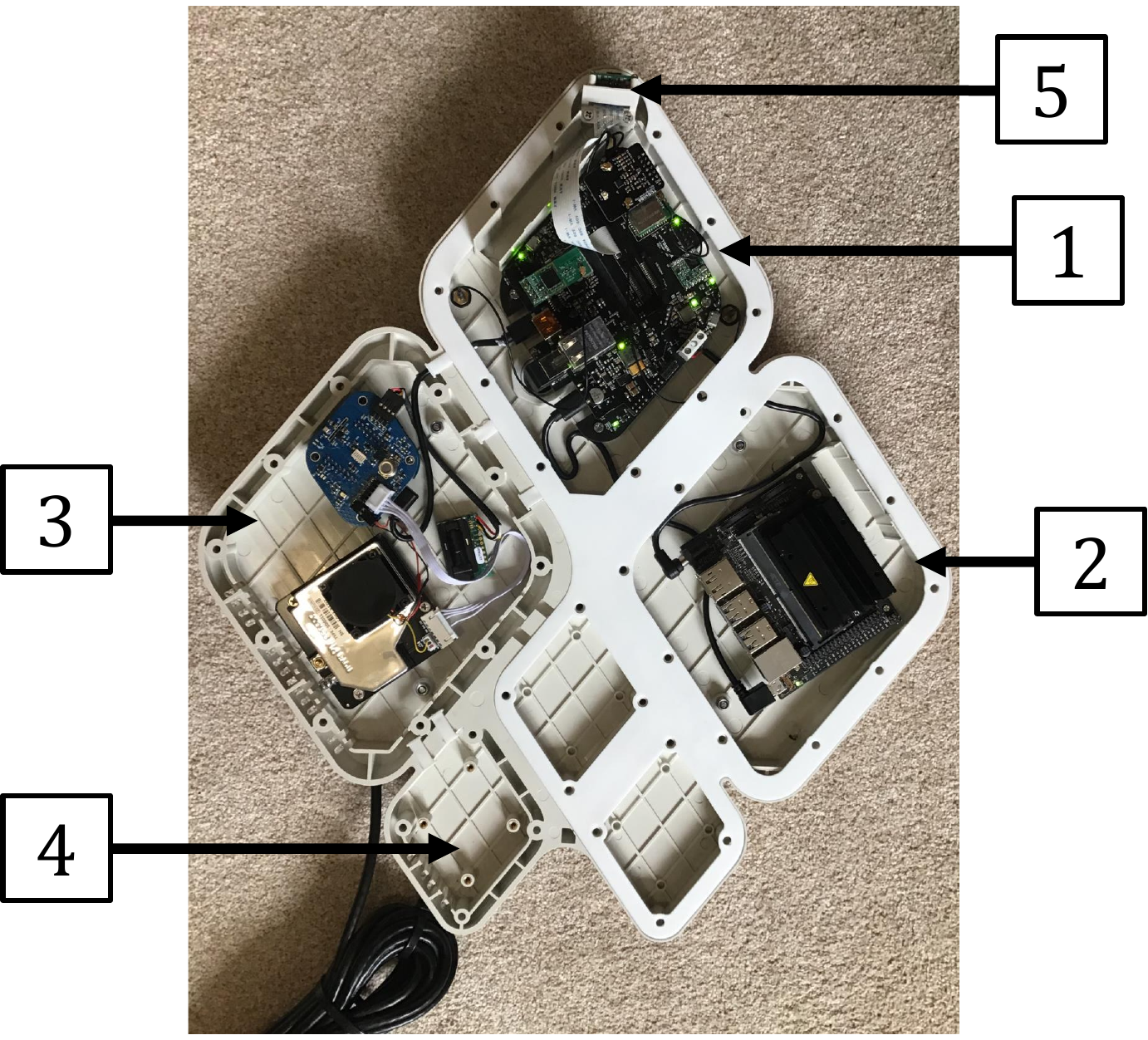}   
      \end{minipage}%
      }
      \hfill
        \subfloat[\label{fig:node connection}]{
      \begin{minipage}[t]{0.45\linewidth}   
        \centering   
        \includegraphics[width=1.6in]{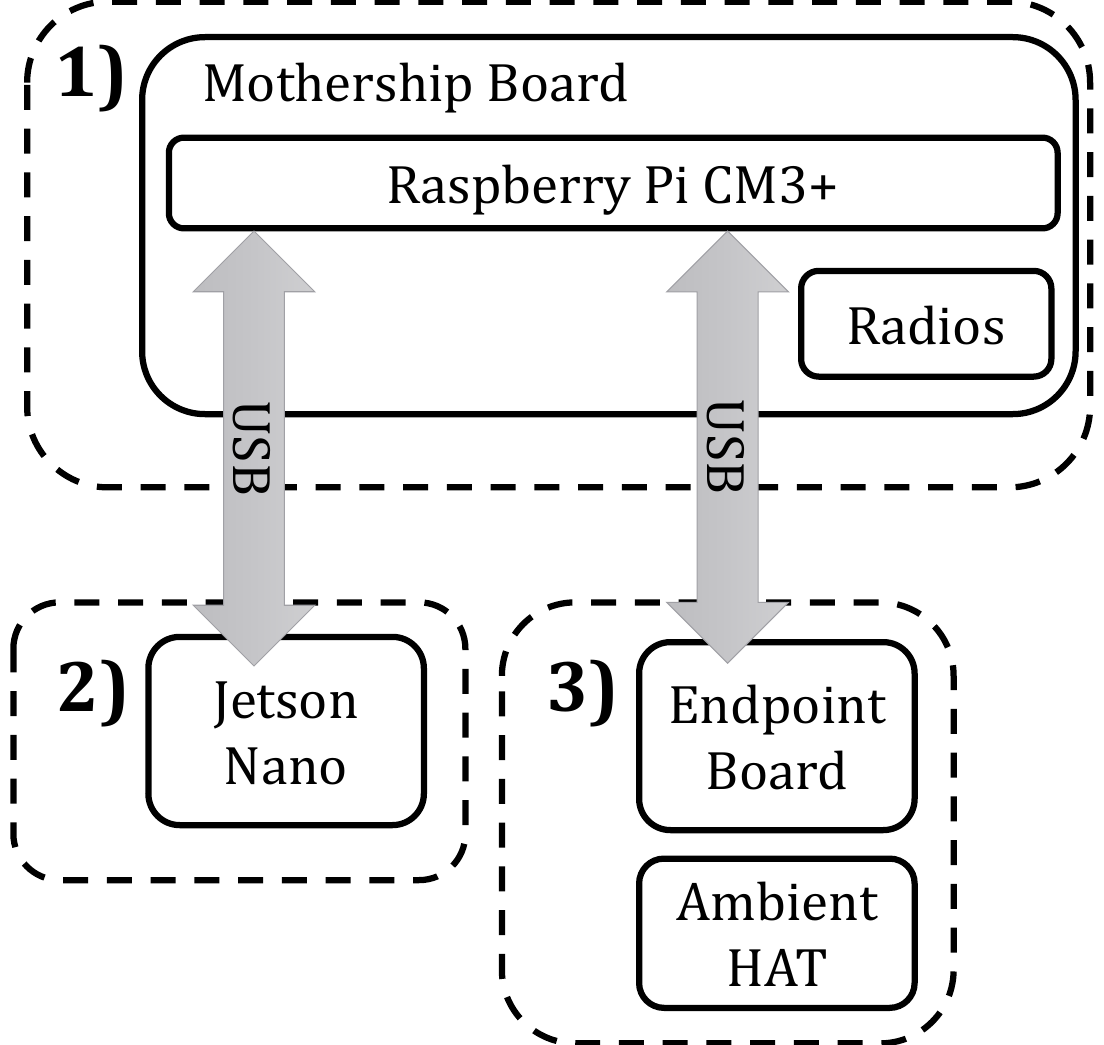}   
      \end{minipage} 
      }
      \caption{(a) shows an \NAME node with the enclosure open, showing modules for 1) mothership, 2) edge, 3) ambient sensing, 4) vacant ambient sensing expansion pod, and 5) the RPi Camera; (b) indicates the connection method of 1) mothership pod, 2) edge computing pod and 3) ambient sensing pod within a node. 
      } \label{fig:node}
\vspace{-1.45mm}
\end{figure}

Fundamentally, the minimum functional unit of \NAME is the \textit{"node"}, which encompasses the mothership pod, edge computing pod, ambient sensing pod, and endpoint boards, as demonstrated in Fig.~\ref{fig:node}. The functionality of each node component is introduced below:
\begin{itemize}[leftmargin=*]
    \item \textbf{Mothership Pod:} It accommodates the Raspberry Pi CM3+ module on a custom carrier board, serving as the operational core of the node. Powered by Raspbian GNU/Linux 10 (buster) 32-bit and a custom kernel (ver. 4.19.95-v7+), the mothership ensures network connectivity and inter-module communication. The carrier board also integrates radio interfaces for network connectivity and wireless testbed capabilities.
    \item \textbf{Edge Computing Pod:} This pod contains an NVIDIA Jetson Nano connected via USB. The Jetson Nano operates on NVIDIA's official Linux4Tegra 64-bit OS (derived from Ubuntu 18.04) with slight modifications. This module enables edge computing and supports GPU-intensive tasks. 
    \item \textbf{Ambient Sensing Pod:} This pod includes an array of environmental sensors. A microcontroller board transmits sensor values to the mothership. The module, housed in a vented enclosure, is exposed to the atmosphere, ensuring accurate readings of the external environment.
    \item \textbf{Endpoint Boards:} The endpoint boards consist of customized Printed Circuit Board Assemblies (PCBAs) centred around a microcontroller, creating an abstraction layer between modular sensors or radios and the node's remaining components. These boards facilitate lower-level communication with sensors through protocols like I2C, SPI, and I2S, effectively offloading time-critical operations from the main processing unit. \NAME nodes offer two Bluetooth and one LoRaWAN interface for end-user access, a Wi-Fi interface for backbone connectivity, and a cellular interface for fail-safe management tasks in the absence of a main LAN connection. 
\end{itemize}

\subsubsection{\NAME Collaborative Robotics Testbed}
The basic element of this collaborative robot (cobot) system is the robot node, which supports multiple sensors, such as infrared, video cameras, Inertial Measurement Units (IMU), temperature, pressure, humidity sensors, etc. Two actuators (Omni-wheels, lift) are mounted in a node. Meanwhile, a number of wireless communication technologies (e.g., Wi-Fi, BLE, UWB, and 5G NR) are integrated into the node. The processor of each node is complemented with GPU (RockPi™ 4B) support, which permits embedded neural network model deployments for on-device processing. The software architecture is based on Docker containers and ROS2 DDS middleware for flexible and extensible evolution to support future sensors or network technologies. The node configuration and telemetry data exchange can be finished through user networks, service networks, access networks, proximity networks, and the meshed network. This testbed can be used for collaborative robotics experimentation, swarm algorithm development and wireless network technology evaluation\footnote{Reader could refer to~ \cite{9369615} for a more detailed description of \NAME cobot testbed.}.

\subsubsection{\NAME Application Implementation}
All experiments and applications are deployed as containers across all devices, orchestrated through a cloud-native platform. A unified backend controls all applications, providing storage and visualisation capabilities and ensuring smooth operation across the entire ecosystem. \NAME users can package their applications as containerised applications, upload them to the \NAME portal, and deploy them on the nodes, where they are orchestrated within multiple Kubernetes clusters. Scheduled experiments operate as stateless instances on either Raspberry Pi or Jetson Nano. 

\section{Existing AI Usecases Under \NAME}
\label{sec:existing AI usecase}
In this section, we present existing example use cases within \NAME, encompassing, i.e., (1) monitoring of street lighting fixtures; (2) digital twin for smart building; (3) large-scale Federated Learning (FL) experiments on \NAME nodes; and (4) intrusion detection for containers on \NAME edge devices. A concise overview figure is illustrated in Fig.~\ref{fig:4ucs}. We briefly describe each use case and discuss how the UMBRELLA ecosystem facilitated their development, delivery and demonstration.

\subsection{Streetlight Monitoring}
\subsubsection{Usecase Description}
This use case involves overseeing the functionality of streetlight fixtures to identify any irregularities or malfunctions promptly. Any unanticipated operations trigger alerts for the council's streetlighting maintenance team. 
Most lights, excluding those following a customised schedule, are programmed to switch off 15 minutes after sunrise and illuminate 15 minutes before sunset. The municipal street lighting team conducts manual checks at regular intervals, approximately every four weeks, to ensure proper functionality. Typically, the team relies on public reports to identify issues with the light fixtures. When multiple fixtures display unexpected behaviour, such as being ON or OFF outside the expected times, the team addresses these issues in batch corrections, implemented over a section of a road for cost efficiency. Our implementation automates this process and provides a more cost-effective way of managing fixed street assets. This use case leverages the deployed \NAME sensor nodes and platform to identify the state of traffic lighting, automates the reporting procedure and reduces the intervals on the scale of days. 
\subsubsection{AI Technique and Results}
The \NAME nodes are equipped with an RPi Camera Module ver.1, facing the streetlight. Taking periodic photos of the fixture, malfunctions can be identified and reported. Our decision-making is based on two methods. We initially employed a simplistic computer vision mechanism, where photos taken from each lamppost are highly exposed. Based on that, it is easy to identify the ON/OFF status. Secondly, an ML classification algorithm was developed using the normal images and providing a confidence interval as well. We have identified that the simple computer vision approach works well at night, while the ML-based approach operates accurately during the day. The ML model operates with an accuracy of $\geq 90\%$ while the computer vision approach with almost $100\%$.

Our use case is still operational and helps the council's team identify malfunctions and abnormalities in the fixtures' operation. Moreover, complementary to this activity we collected and shared a dataset with the research community~\cite{mavromatis2022dataset}. The dataset was automatically labelled based on the mechanisms mentioned above and comprised 350,000 images, taken over a six-month period using 140 UMBRELLA nodes.

\begin{figure}[t]   
    \subfloat[\label{fig:uc1}]{
      \begin{minipage}[t]{0.45\linewidth}
        \centering 
        \includegraphics[width=1.7in]{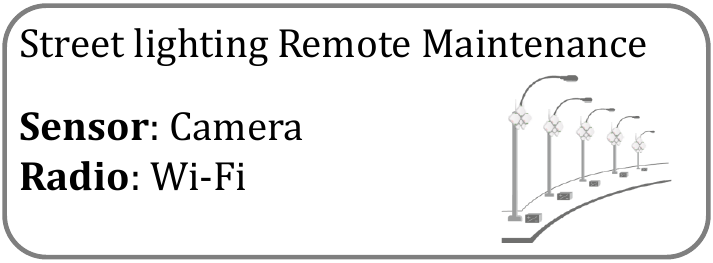}   
      \end{minipage}%
      }
      \hfill
        \subfloat[\label{fig:uc2}]{
      \begin{minipage}[t]{0.45\linewidth}   
        \centering   
        \includegraphics[width=1.7in]{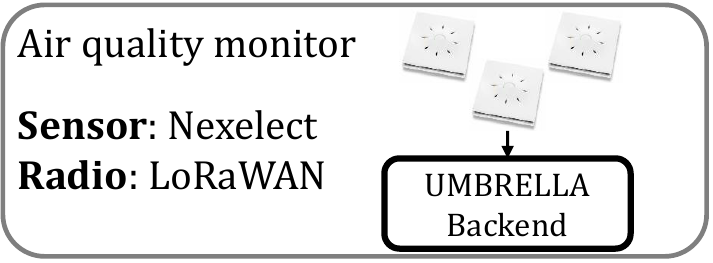}   
      \end{minipage} 
      }\\
      \subfloat[\label{fig:uc3}]{
      \begin{minipage}[t]{0.45\linewidth}
        \centering 
        \includegraphics[width=1.7in]{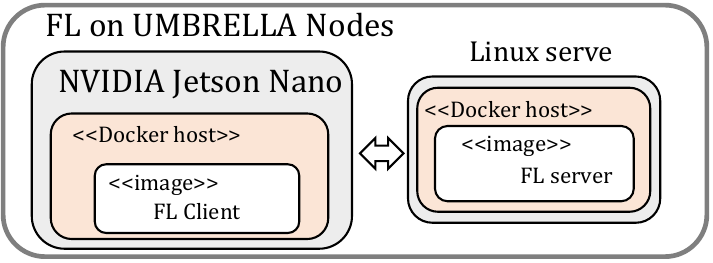}   
      \end{minipage}%
      }
      \hfill
        \subfloat[\label{fig:uc4}]{
      \begin{minipage}[t]{0.45\linewidth}   
        \centering   
        \includegraphics[width=1.7in]{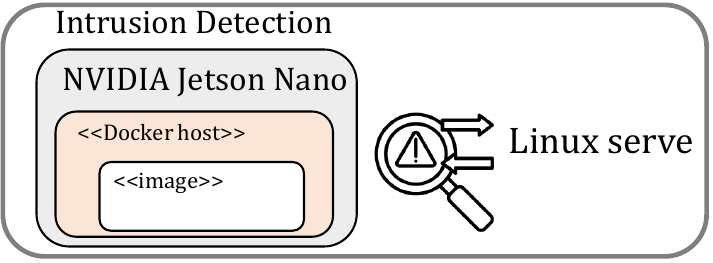}   
      \end{minipage} 
      }
      \caption{This figure illustrates an overview of the \NAME AI use cases presented in this paper, where (a) refers to the street lighting remote monitoring; (b) is digital twin for building environment sensing; (c) depicts the FL trail on \NAME node; (d) shows the intrusion detection on \NAME edge devices.
      } \label{fig:4ucs}
\vspace{-1.25mm}
\end{figure} 
\subsection{Digital Twin for Buildings' Environmental Sensing}
\subsubsection{Usecase Description}
In the wireless front, \NAME integrates various technologies to facilitate short- and long-range communications. Bluetooth and LoRaWAN particularly are accessible by end users and be used for experimentation. This use case focuses on temperature and air quality (AQ) monitoring, from sensor data collected from devices remotely connected to \NAME. The data collected are later stored in \NAME's backend and processed in real-time. Some examples of real-time sensor data collected include temperature, CO2, humidity, pressure, noise, etc.

\subsubsection{AI Technique and Results}
The ultimate goal of this use case is to construct a high-fidelity 3D data-driven digital twin that virtually ``senses'' a building environment. In other words, based on a small scale of measurements, can extrapolate the expected air quality in the future. In order to expedite the rendering of the virtual environment and minimise the cost of sensor anchor points (AP), an autoencoder-based ML model is utilised. This model is responsible for generating virtual APs and inferring their measurements from the real-time sensor data stream deployed within the same experimental setting while preserving fidelity.

For this use case, remote nodes connected wirelessly to the \NAME infrastructure were used. These nodes are battery-powered small-factor sensor boards. The sensor APs are linked to \NAME's platform, transmitting measurements every 10 minutes. \NAME facilitated all the computing resources required for model training and inference as well as the infrastructure for storing and visualising the data. The sensor boards were installed in care homes in the South Gloucestershire Council. Remarkably, the autoencoder model developed achieves an absolute percentage error of less than 1\% for temperature and 10\% for CO2 measurements when compared to reference sensors while reducing the number of deployed sensors by one-fourth. The detailed description of this use case is available in~\cite{buildtwin2023}.

\subsection{Large-scale FL Experimentation}
\label{subsec:FL Experiments with FashionMNIST}
\subsubsection{Usecase Description}

\begin{figure}
\centering
\includegraphics[width=0.8\linewidth]{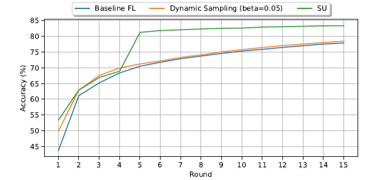}
\caption{Preliminary results using four connected \NAME IoT devices and containerized FL system.}
\label{fig:fl_results}
\vspace{-1.25mm}
\end{figure}

Federated Learning (FL) is a decentralised machine learning framework designed to parallelise training across multiple interconnected devices simultaneously training a single model. In the context of \NAME, UMBRELLA edge nodes, serve as network entry points and continuously collect raw data. In certain scenarios, IoT devices may face limitations in network quality, impacting their ability to share model parameters for FL. Transmitting substantial data volumes through low-power resource-constrained IoT networks (e.g. Bluetooth or LoRaWAN) becomes a significant bottleneck in FL. Some devices may also have constraints on energy usage, communication bandwidth, and other network resources, hindering their participation in model training. 
This use case aims to develop an FL system capable of training on \NAME edge devices while minimising the transmission of model parameters. This approach offers the dual benefit of reducing network bandwidth requirements, improving quality of service, and lowering energy consumption during training. Additionally, the nature of FL ensures that the entire training dataset (raw sensor data) remains confidential, maintaining privacy during FL training.
\subsubsection{AI technique and Results}
In this large-scale FL experiment, levering the flexible configuration capability of \NAME, we implemented a Selective Updates (SU) FL method~\cite{anwar2022system} to minimise communication volume in FL training while maintaining inference performance. \NAME's distributed edge processing nodes were used for scaling up our experiment and evaluating our scenario across a realistic system. The proposed method operates on a client-round basis, employing a sliding/adaptive observation window and threshold to effectively gather local client model performance before deciding whether to update our Parameter Server (PS) or conduct further local training. The model deployment and update procedure are illustrated in Fig.~\ref{fig:uc3}. We utilised the MNIST Fashion dataset~\cite{xiao2017fashion} for testing, where each image is $28\times28$ pixels and greyscale, featuring ten classes for the FL model to detect and classify. The results, presented in Fig.~\ref{fig:fl_results}, within \NAME platform, the SU method has shown substantial advantages in terms of convergence speed and accuracy compared to baseline FL and dynamic sampling.

\subsection{Intrusion Detection System at IoT Edge}
\label{subsec:Intrusion Detection System at IoT Edge}
\subsubsection{Usecase Description}
The open-access nature of \NAME, where end-users can deploy their containerised experiments raises security concerns for the functionality of the system. Attacks like privilege escalation, can compromise both the host and the underlying IoT infrastructure. Inspired by that, we developed a semi-supervised FL solution, aiming to detect such issues at the edges, and ensured the uninterrupted operation of the \NAME platform. The ultimate goal of this use case is to train and deploy a real-time anomaly detection system with updated models throughout the entire \NAME IoT infrastructure.

\subsubsection{AI Technique and Results}
A novel systematic approach is proposed to perform intrusion detection on the \NAME edge, to detect any malicious activity by a container. The approach is divided into two parts: data pipeline (ingesting the data from multiple hosts) and modelling (training model from ingested data). The data pipeline begins by processing raw auditd logs at the edge, extracting field names and values. Then, event-based feature engineering is applied based on a methodology validated in~\cite{tien2019kubanomaly}. This streamlined pipeline transforms raw logs into a structured and enriched dataset for efficient analysis and insights. In the modelling phase, three scenarios are considered for generating training/test datasets: $S_A$, where container escape involves launching a denial of service (DoS) attack using a host shell; $S_B$, where container escape involves writing to host permission files, granting passwordless sudo permission to the user; and $S_C$, a benign container with an Apache PHP webserver without any escape~\footnote{The dataset of container escape detection is available on~\cite{pope2021dataset}.}. 
In each dataset, there are three classes of labels for intrusion detection, including the \textit{"Normal"}, \textit{"Anomaly"}, and \textit{"Uncertainty"}. The $S_C$ dataset is split for the pre-training and validation of the autoencoder model, which is subsequently validated on the $S_A$ and $S_B$ datasets. After pre-training the autoencoder, the model undergoes FL updates. As in the previous use case, the \NAME edge computing nodes and backend provide the basis and underlying infrastructure for the FL deployment. In this case, the FL training follows a classical manner. That is, in each FL round, the model is independently trained on individual clients, and the FL model is aggregated to the server to update the global model. This aggregated model is then broadcast to each client for the next FL round. The final FL model can reach more than 98\% of identifying accuracy on $S_A$ and $S_B$ datasets.

\section{Future AI Usecases Towards Crowdsensing}
\label{sec:future AI}
In the future, it is expected that \NAME will serve as a distinctive platform for exploring and advancing IoT/IIoT technologies. Due to \NAME's nature and its real-world deployment, Crowdsensing applications and demonstrations can leverage the existing capabilities. \NAME nodes installed across a very busy road in the northern part of Bristol, and inside the University of West of England campus can serve as the underlying infrastructure for real-world Smart City Crowdsensing applications. 
Additionally, through the utilisation of its cobot testbed, inventive mechanisms can be evaluated enabling robot-to-robot and robot-to-human interactions. This involves enhancing data exchange capabilities, improving collaborative abilities, and ultimately enhancing overall efficiency in co-working. These representative examples are illustrated as follows. 

\subsection{Smart City \& \NAME AI}
Smart Cities are defined as an urban environment that leverages technology and data-driven solutions to improve efficiency, sustainability, and the quality of life for its residents. 
Undoubtedly, AI plays a crucial role in achieving these objectives. The \NAME system is anticipated to contribute significantly to the future smart city landscape. \NAME's comprehensive sensing, computing, and robust AI capabilities form the basis for developing reliable crowdsensing systems for smart cities. These systems utilise sensor data as a criterion for intelligent decision-making, encompassing tasks such as air pollution monitoring and traffic control. Moreover, the edge computing device on the \NAME node can seamlessly function as a complimentary multi-access edge computing platform. This functionality facilitates the smooth offloading and balancing of cloud computing demands from pervasive computing, particularly in domains like environmental monitoring, video surveillance, and the evolution of intelligent transportation systems.

\subsection{Robotics \& \NAME AI}
The \NAME cobot testbed is designed for experiments where multiple robots work together or jointly with humans for a shared goal. Cobot can naturally be seen as a crowdsensing system. The success of the cobot task heavily relies on effective communication (data exchange) among them to process task allocation and coordination, sensor data sharing, position and location sharing, status updates, error detection, and recovery. 
Reducing the latency of such data sharing and decision-making will be the focus of the next step of \NAME cobot-based crowdsensing system development. From the techniques perspective, potential consideration will be given to the following points.
\begin{itemize}[leftmargin=*]
    \item \textbf{Semantic communication:} It provides a promising technique to compress such raw sensor data in a more meaningful and context-aware manner. Deploying the embedded semantic encoder/decoder model on the \NAME robot node can reduce the communication overhead, and improve the efficiency of cooperative robotics systems~\cite{li2023open}.
    \item \textbf{Communication and control co-design:} This approach entails the collaborative design and optimisation of communication and control systems, with the former playing a crucial role in transmitting control information and feedback signals among system components. This co-design methodology is anticipated to enhance the robustness, agility, and determinism of the cobot system. By customising communication protocols on \NAME node to meet the specific requirements of control algorithms, we can minimise unnecessary data transmission and reduce processing overhead.
    \item \textbf{Automated Planning:} The completion of this process is expected to be achieved by harnessing the inherent AI capabilities of the \NAME robot node and utilising connected cloud/edge intelligence. This will enable the automatic determination of optimal paths and collaborative methods for the involved robots, facilitating a streamlined operation for the entire cobot system.
\end{itemize}

\section{The need for full-scale MLOps}
\label{sec:MLOps}
MLOps is a method of controlling the lifecycle of an ML model, right from design to deployment. It encompasses a set of principles that merge ML, DevOps, and data engineering to ensure the dependable and efficient deployment and maintenance of ML models in a production environment. It essentially involves delivering ML applications through the lens of DevOps, with a particular emphasis on data and ML models. MLOps is all about streamlining and expediting processes, wherein the streamlining entails automating the ML pipeline, covering the journey from data processing to model development for continuous training, as well as automating the Continuous Integration/Continuous Deployment (CI/CD) processes for ML applications~\cite{9931127}. Expediting refers to the ability to enhance the speed of delivering these applications while upholding service quality standards. The practice of the MLOps system brings three benefits for the ML model development and deployment, including (1) automated and effective model deployment via a unified workflow; (2) easy for quality control, ensuring the reliability of ML models in the real-world; (3) speed up model management and updating. 

So far, there is a missing native MLOps platform that is adopted or specifically designed for \NAME. The implementation and operation of applications in \NAME are predominantly handled by Kubernetes orchestration systems, functioning as a DevOps platform. Consequently, the capabilities of \NAME in terms of ML-related automation functions, such as processing training data, conducting model training/validation and ML model lifecycle management, are relatively limited. This limitation could pose challenges to the establishment of trustworthy crowd-sensing systems. Consequently, we intend to delve into the creation of an in-\NAME MLOps platform. It is anticipated that all the AI use cases outlined in this paper will be built upon the native MLOps infrastructure in the future.



\section{Conclusions}
\label{sec:discussion and conclusion}
This paper provides a comprehensive overview of the AI/ML applications crafted within the \NAME framework. Four use cases accompanied by in-depth technical details and the accomplished outcomes, are presented. Simultaneously, this paper explores potential crowdsensing applications in smart cities and robotics that are earmarked for future development, underscoring the potential growth of \NAME. Additionally, the advantages of embracing full-scale MLOps in the context of \NAME are highlighted. The authors humbly hope that this paper can positively contribute to the advancements of both the theoretical understanding and practical implementation of intelligent IoT application development.

\section*{Acknowledgments}
This work was supported by Toshiba Europe Ltd. and Bristol Research and Innovation Laboratory (BRIL).

\bibliographystyle{IEEEtran}
\bibliography{references}
\clearpage

\end{document}